\documentclass{article}

\usepackage[pdftex]{graphicx}  
\usepackage{lineno,hyperref}
\modulolinenumbers[5]

\usepackage{gensymb}
\usepackage{amsmath}
\usepackage{caption}
\usepackage{subcaption}
\usepackage{authblk}

\title{ The new HADES ToF Forward Detector}
\author[1]{A. Blanco}
\author[1,2]{P. Fonte}
\author[1]{L. Lopes}
\author[1]{J. Saraiva}
\affil[ ]{for the HADES collaboration}
\affil[1]{Laboratory of Instrumentation and Experimental Particles Physics, Coimbra, Portugal}
\affil[2]{Coimbra Polytechnic - ISEC, Coimbra, Portugal}

\begin{document}
	\maketitle

\begin{abstract}

The High-Acceptance DiElectron Spectrometer (HADES) at GSI Darmstadt consists of a 6-coil toroidal magnet centered on the beam axis and six identical detection sections located between the coils and covering polar angles between $18^\circ$ and $85^\circ$. The physics aims include the study of the properties of hot and dense hadronic matter as well as elementary and pion-induced reactions.

To increase the acceptance of HADES at very low polar angles in the forward region, between $0.5^\circ$ and $7^\circ$, a new detector, the Forward Detector (FD), has been built. The FD is composed of a tracking and a Time Of Flight (TOF) detector based on Resistive Plate Chamber (RPC) technology. The TOF detector, covering an area of around $2$~m$^2$, is composed by $128$ strip-like shielded RPC cells, with two different widths $22$~mm and $44$~mm and $750$~mm length distributed in four modules symmetrically placed around the beam axis. Each cell is composed by four gas gaps, $0.270$~mm, delimited by three ($2$~mm) aluminum and two ($1$~mm) glass electrodes. In order to cope with an expected maximum particle load of around $400$~Hz/cm$^{2}$, close to the beam axis, the detector is operated above room temperature in order to decrease the resistivity of the glass and increase the count rate capability.

Details of the system construction and results concerning timing precision are described in this communication. The detector was operated at $31.5^\circ$C with a maximum particle load of around $600$~Hz/cm$^2$ during a production beam time for six weeks in early $2022$ showing an average time precision of around $80$~ps.

\end{abstract}

\section{Introduction}
The HADES (High-Acceptance DiElectron Spectrometer) at GSI Darmstadt is a fixed target experiment  consisting of a 6-coil toroidal magnet centered around the beam axis and six identical detection sections, sectors, located between the coils covering polar angles between 18° and 85° \cite{Had09}. Each sector is equipped with the following subsystems. A Ring-Imaging Cherenkov (RICH) detector used for e$^{+}$ e$^{-}$ identification. Four Mini-Drift Chambers (MDCs), two in front of and two behind the magnetic field, for particle momentum reconstruction based on the deflection angle of particle trajectories derived from the hit positions before and after magenetic field. Two detectors, based on scintillators and Resistive Plate Chambers (RPCs) technology, for time-of-flight (TOF) measurements in combination with a diamond start-detector located in front of a segmented target. An electromagnetic calorimeter (ECAL) for photon detection. Additionally, the set-up is completed by  a forward hodoscope used for event plane determination.

The physics aims include the study of the properties of hot and dense hadronic matter—a key problem in heavy-ion physics—as well as elementary and pion-induced reactions in the few GeV energy regime. A summary of recent physics results can be found in \cite{Had19b}.

To increase the acceptance of HADES at very low polar angles in the forward region, between $0.5^\circ$ and $7^\circ$, a new detector, the Forward Detector (FD), has been built. The aim of the FD is to measure the electromagnetic decays of the hyperon resonances, as well as the production of double strange baryon systems in p + p reactions at a beam kinetic energy of $4.5$~GeV \cite{Had21}.

The FD is composed of a tracking detector, based on straw tubes technology \cite{Str18} and a TOF detector based on RPCs, the fRPC-TOF. The requirements for the TOF detector are an efficiency $> 90$\%, a time precision $< 100$~ps and a maximum counting rate (close to the beam axis) of about $400$~Hz/cm$^2$. To shorten the development time of the fRPC-TOF, it was decided to use exactly the same technology used in the former HADES RPC-TOF detector \cite{HADES}. However, to cope with the expected maximum particle load the detector was constructed with thinner glass electrodes and is operated above room temperature to decrease the resistivity of the glass and increase the count rate capability (one order of magnitude reduction on resistivity is expected every $25^\circ$C \cite{Gon05}). This concept was previously tested in a small prototype \cite{Bla22} and  results suggest that the count rate capability can be extended at least up to $1500$~Hz/cm$^{2}$ when the detector is operated at $40.6^\circ$C without noticeable loss of efficiency or timing precision degradation.

This article describes the main parts of the fRPC-TOF in its final configuration after installation in the HADES spectrometer. Moreover, first results showing an average time precision of around $80$~ps when the detector is operated at $31.5^\circ$C with a maximum particle load of around $600$~Hz/cm$^2$ during a production beam time for six weeks in early $2022$, are also shown. 

\section{The forward RPC TOF}
\subsection{The RPC}
\label{subsec:}

\begin{figure} [t]
	\centering 
	\includegraphics[width=\linewidth]{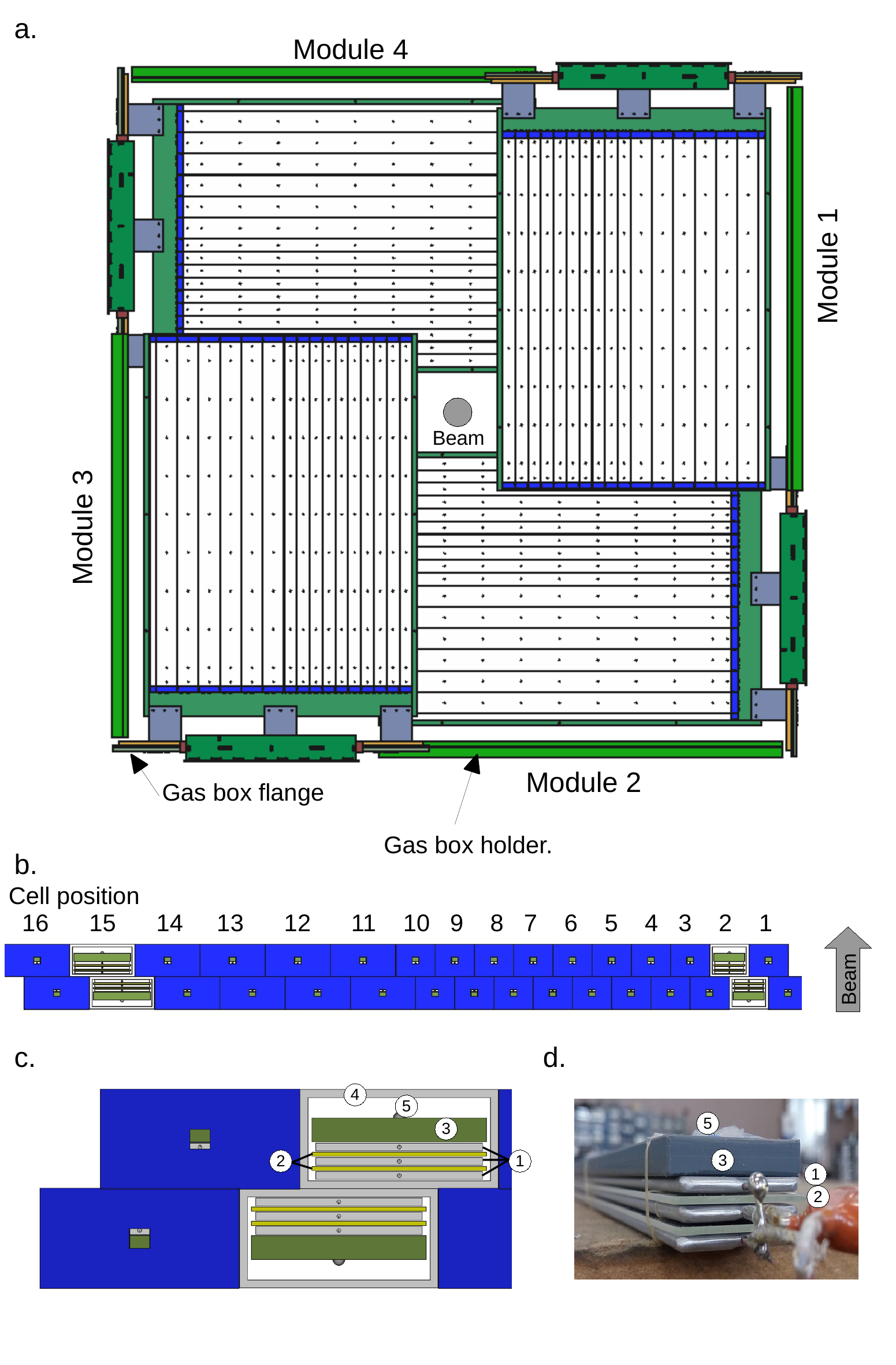}
	\caption{a. Arrangement of the modules around the beam axis (gas box not shown) from the target point of view. b.  Cross-section of a module showing the distribution of 32 cells in two layers. Twenty 22 mm wide cells near the beam axis and twelve 44 mm wide cells in the peripheral part. c. CAD detail of two of the cells inside the aluminum tube without end-shields. d. Actual photograph of one of the cells showing: 1- aluminum electrodes, 2- glass electrodes, 3- PVC pressure plate, 4- 2 mm thick aluminum shielding tube and 5- compression springs.}
	\label{fig:frpcSchematic}
\end{figure}

A circular area of approximately $1300$~mm diameter (polar angle $<7^\circ$) is covered by four equal modules positioned around the beam axis at every $90$~$^\circ$ as shown in \ref{fig:frpcSchematic}.a, leaving uncovered a square of about $200$~mm side to allow the beam to pass through. Modules $1$ and $3$ are in the same plane at a distance of $5700$~mm from the target while modules $2$ and $4$ are at a distance of $5850$~mm from the target. There is a small overlapping area between all modules to avoid dead areas and to facilitate calibration. 

Each module is composed of $32$ individually shielded \cite{GRPC} strip-like tRPC cells \cite{HADES} distributed in two layers, sixteen cells in each layer, as shown in \ref{fig:frpcSchematic}.b. Each individual cell consists of three aluminum, $2$~mm thick, and two glass electrodes\footnote{Soda-lime with a bulk resistivity of around $4$x$10^{12}$~$\Omega$cm at $25$~$^\circ$C.}, $1$~mm thick, with a length of $750$~mm and two different widths $22$~mm and $44$~mm. All of the electrodes have rounded edges and the glass exceeds the aluminum in size by $1$~mm (in both dimensions) to prevent discharges, see figures \ref{fig:frpcSchematic}.c and \ref{fig:frpcSchematic}.d. The gap is defined by PEEK (Polyetheretherketone) mono-filaments of $0.270$~mm diameter, spaced approximately by $100$~mm along the cell. The stack is housed inside individual aluminum tubes (shielding) and compressed by springs on top of each mono-filament that apply a controlled force through a PVC (Polyvinyl chloride) plate that distributes the force.

High-voltage (HV) close to $6$~kV is applied to the central aluminum electrode via $1$~M$\Omega$ resistors and high voltage cables, while the outer electrodes are grounded and the glass electrodes are kept electrically floating. Insulation to the shielding tube walls is assured by a triple-layer KAPTON\texttrademark\hspace{0.1cm} adhesive laminate. An end-shield made of aluminum foil is glued to both ends of the aluminum tube to produce a fully shielded element. The signals are collected, at both ends of each chamber, by coaxial cables through $2$~nF HV coupling capacitors and extracted from the gas box through a flange via Radio Frequency Micro Miniature CoaXial (MMCX) connectors. 

The two different cell widths in each module, twenty $22$~mm wide cells close to the beam and twelve $44$~mm wide cells at the periphery, are aimed at accommodating the particle flux (larger near the beam axis) and thus keeping the rate per cell ($\approx 30$~kHz) and the pile-up ($\approx 0.5$\%) relatively constant as a function of the radial direction. The two layers within each module, shifted half the width of the cell, are aimed at creating a surface without efficiency gaps, one layer covers the efficiency gaps created by the other layer due to the insertion of the aluminum tubes.

The two layers are packaged together and placed inside a gas box made of 2 mm thick aluminum sheet with only one flange at one side, see figure \ref{fig:frpcSchematic}.a. It is in this flange that the layers are mechanically secured and where all the feed-throughs are installed: HV, gas and signal. The gas box is fixed on one side by a thick aluminium rod, gas box holder in figure \ref{fig:frpcSchematic}.a., and held in place with the help of a gantry made of a technical profile that ensures the correct positioning of the modules around the beam axis, see figure \ref{fig:frpcSchematic}.a. The gantry stands on a trolley, made of the same technical profile, to lift the four modules to the correct height of the beam axis.

\subsection{Readout}
\label{subsec:readout}
The entire readout system used in the fRPC-TOF is similar to that used in the RPC-TOF of HADES \cite{HADES}. The signals created by each cell (two, one at each end) are amplified and discriminated on four-channel Daughter Boards (DB) \cite{HADES_FEE} Front End Electronics (FEE). These are able to encode in a single output signal: the time (leading edge), with a precision of $< 35$~ps $\sigma$, and the charge (pulse width). The charge is obtained by measuring the Time over Threshold (ToT) on a copy of the amplified signal, integrated with an integration constant of approximately $100$~ns. Each of the layers within a module is read with eight DBs supported on a Mother Board (MB) \cite{HADES_MBDB} providing: mechanical support, power and communication (thresholds settings). The low voltage system that powers this electronics is based on a custom made distributed low voltage system \cite{HADES_LV}. The resulting signals are read out by the TDC-Readout-Board (TRB)\cite{TRB3} equipped with $128$ multi-hit Time to Digital (TDCs) (TDC-in-FPGA technology) channels with a time precision better than $20$~ps $\sigma$. The reading of the entire fRPC-TOF	, comprising $256$ channels, is performed by two TRBs.

\subsection{Heating system}
\label{subsec:frpcPhoto}

\begin{figure}
	\centering 
	\includegraphics[width=\linewidth]{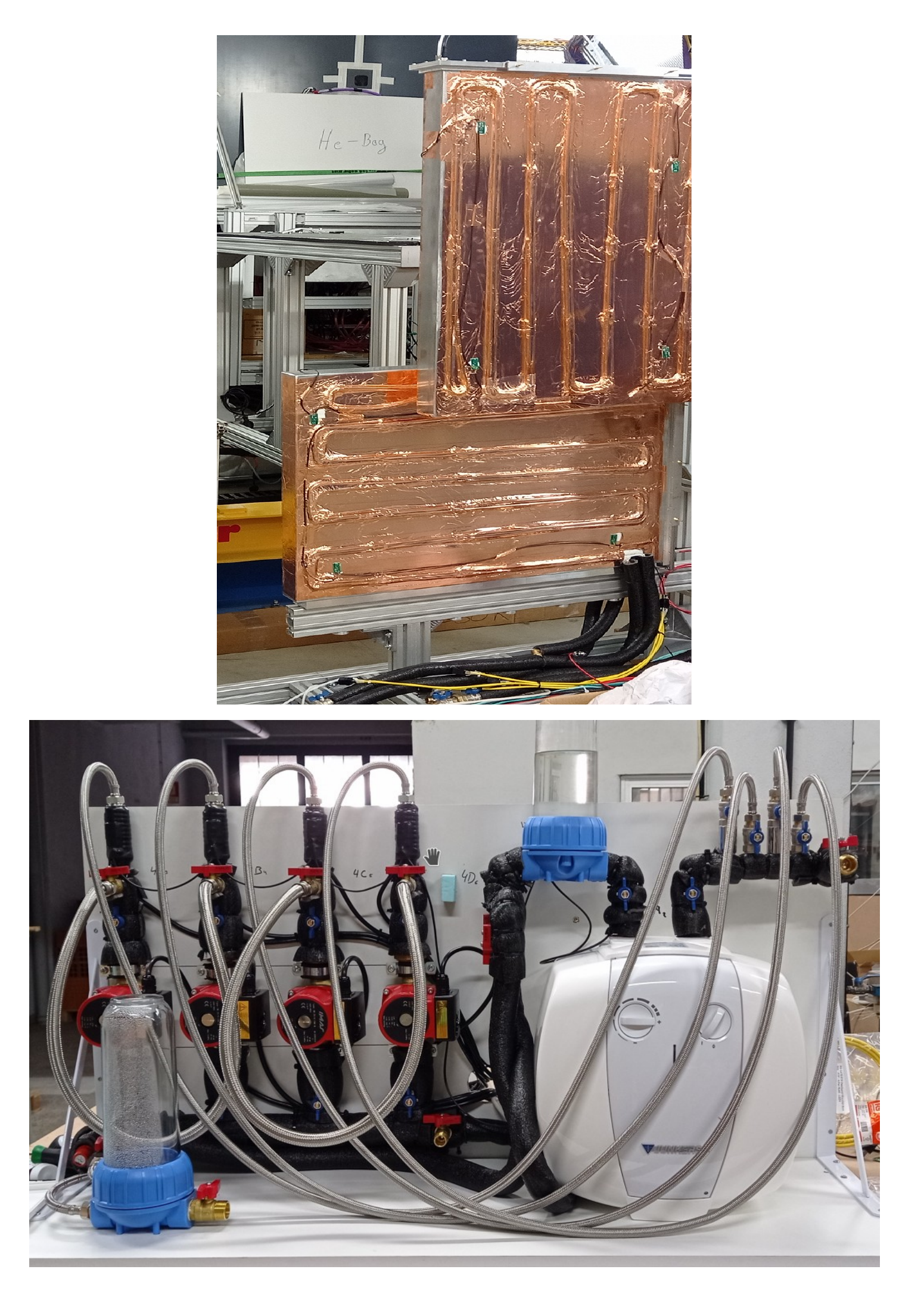}
	\caption{a. Detailed view of the radiator installed in two of the modules already placed on the gantry. The copper pipe under the copper tape and some of the temperature sensors (small green printed circuit boards) are visible. The black tubes are the insulated inlet and outlet pipes. b. Shared heating element and pumps before installing the XPS insulation.}
	\label{fig:heating}
\end{figure}

The heating system of the fRPC-TOF is based on a hot water circulation system. With this system, heat can be provided to the coldest areas or extracted from the hottest areas, such as the FEE electronics (the FEE need to read a module dissipates $40$~W).

\begin{figure}
	\centering 
	\includegraphics[width=\linewidth]{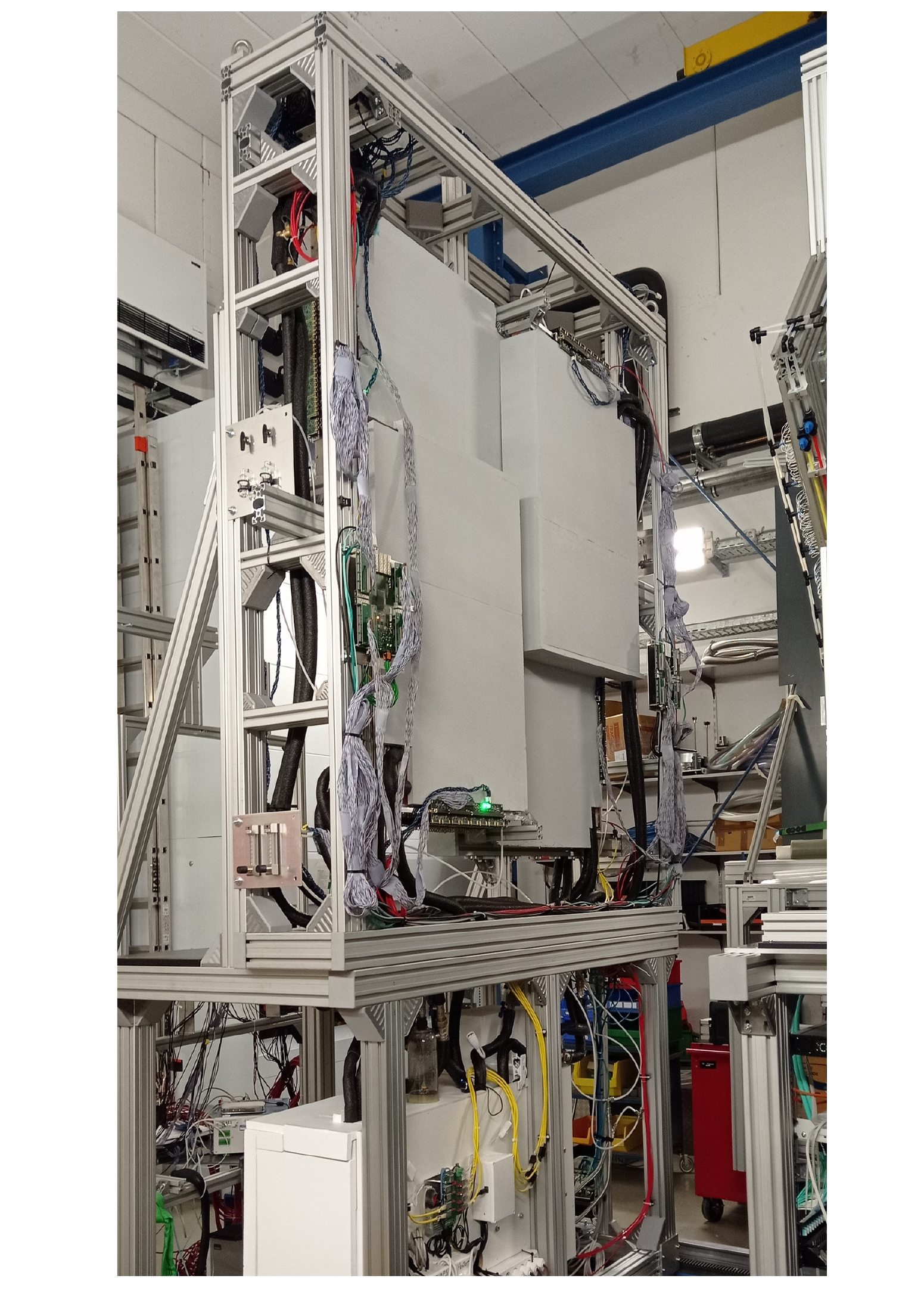}
	\caption{fRPC installed at the forward of the HADES spectrometer. On the lower trolley is installed the heating and pump system (white box) and the communication and power supply systems. On the upper gantry are the four modules with the XPS insulation together with the FEE and the two TRB boards.}
	\label{fig:frpcPhoto}
\end{figure}

The system consists of a radiator and a heating and pumping system. The radiator, see figure \ref{fig:heating}.a, consists of a $10$~mm copper pipe that snakes over each of the modules (on both sides) making the same inlet and outlet route. The pipe is glued to the module with thermal glue and copper tape to facilitate heat transfer to the module. The heating system, see figure \ref{fig:heating}.b, consists of a shared heating element with $1500$~W of power to which four pumps are connected to circulate the water in each of the modules. The two parts are connected with domestic fittings used in trunking. The whole assembly, modules and connections, are insulated with XPS foam to isolate it from the environment, see figure \ref{fig:frpcPhoto}.

The monitoring system has 8 temperature sensors (I2C TMP75) on the outside of each module and another 8 on the inside. A Proportional-Integral-Derivative algorithm controls the heating system to keep the temperature inside the modules constant. This is kept stable within a range of $\pm 0.5^\circ$C (difference from maximum to minimum) with a power consumption of $210$~W and $450$~W with a thermal differential for the environment of $14^\circ$C and $22^\circ$C respectively. 

The HV system is based on a CAEN A1526, 6 Ch 15 kV Common Floating Return HV module (only four channels, one per module, are used) mounted in a CAEN mainframe SY4527. The detector is operated in open gas loop with a mixture of $97$\%  C$_{2}$H$_{2}$F$_{4}$ and $3$\% SF$_{6}$.

Figure \ref{fig:frpcPhoto} shows the fRPC installed at the forward of the HADES spectrometer. On the lower trolley is installed the heating and pump system (white box) and the communication and power supply systems. On the upper gantry are the four modules with the XPS insulation together with the FEE and the two TRB boards.

\section{Results}

\begin{figure}
	\centering 
	\includegraphics[width=\linewidth]{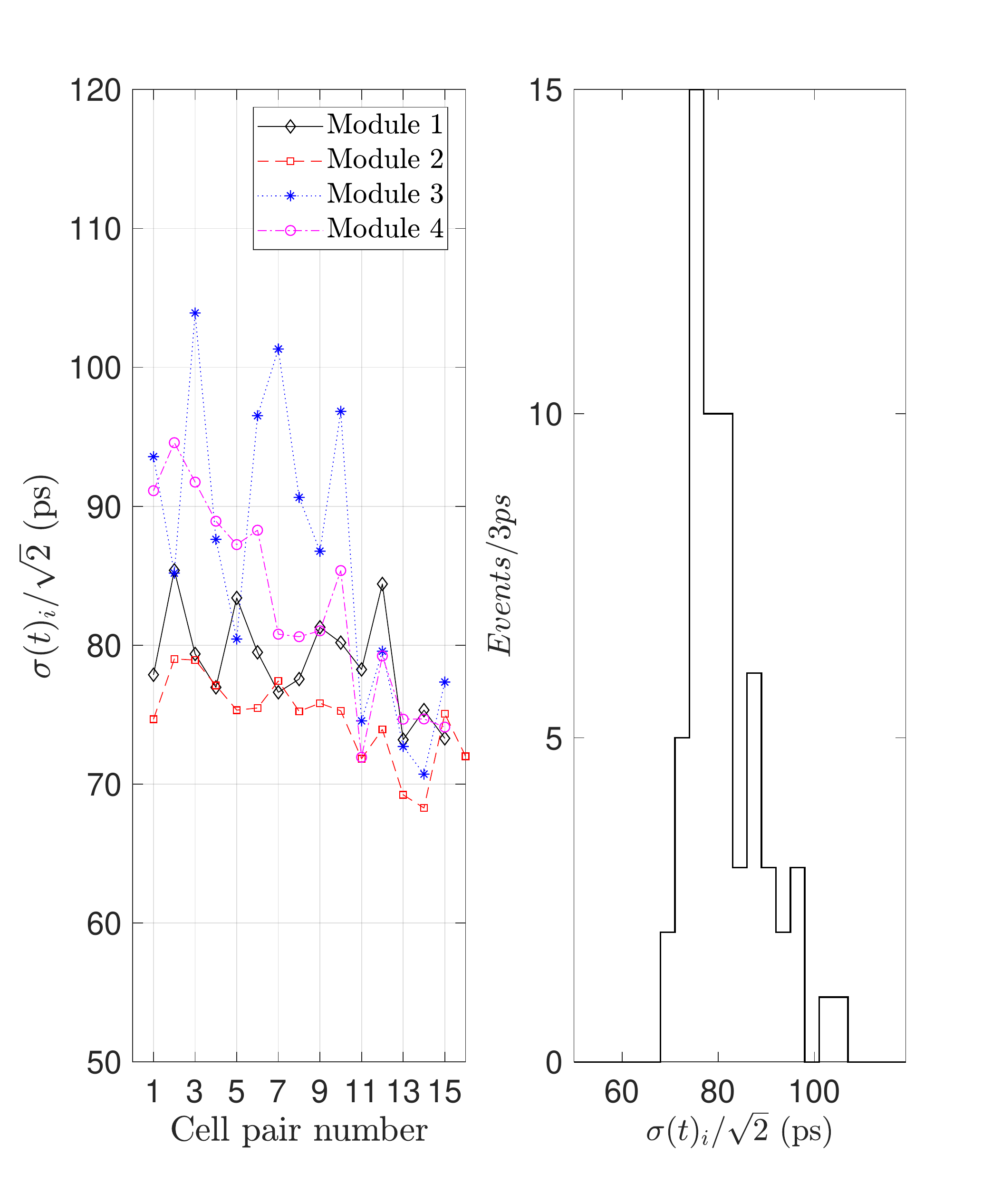}
	\caption{a. $\sigma(t)_i/\sqrt2$, sigma of the time difference between two cells of different layers at the same position within the layer divided by $\sqrt2$, as a function of the cell pair number for the four position. b. Histogram of all $\sigma(t)_i/\sqrt2$ with an average value of $\approx 80$~ps}
	\label{fig:results}
\end{figure}

The fRPC-TOF was exposed for six weeks to collisions of a $4.5$~GeV proton beam on a liquid hydrogen target (p + p collisions). The beam intensity was around $8$x$10^{7}$~protons/second and the spill and duty cycle were $20$ seconds and $95$\% respectively.

Under these conditions the fRPC-TOF was exposed to a maximum particle rate, in the region close to the beam axis, of approximately $600$~Hz/cm$^{2}$ and a factor four less in the peripheral part. The detector was operated at a temperature of $31.5^\circ$C with a reduced electric field of $400$~Td, which corresponds to a voltage of $2700$~ Volts/gap.

Thanks to the overlap between the two layers of each module, it is possible to calculate the intrinsic time precision of each pair of cells, $\sigma(t)_i$, where $i$ is a certain pair of overlapped cells. To calculate this parameter, the time difference between two cells of different layers at the same position within the layer is calculated for each of the modules. This time difference is corrected as a function of  charge, slewing correction, and as a function of the position inside the cell, which tries to correct the observed dependence of the average time along the detector due to mechanical imperfections. In order to characterize the resulting non Gaussian distribution, the $\sigma$ of a Gaussian fit within $\pm1.5$~$\sigma$ about the mean of the original distribution was calculated. Assuming equal precision for both cells, the intrinsic time precision for a single cell is the calculated value divided by $\sqrt2$.

Figure \ref{fig:results}.a shows $\sigma(t)_i/\sqrt2$ as a function of the cell pair position in the module for the four modules. The time precision is slightly worse for smaller cell positions (close to the beam axis). This dependence is under study and may be a consequence of the particle load or the energy of the particles. Figure \ref{fig:results}.b shows the histogram of all $\sigma(t)_i/\sqrt2$ with an average value of $\approx 80$~ps.

\section{Conclusions}
A new TOF detector for the forward region of the HADES experiment based on RPCs has been built and installed in the experiment. The detector is composed of 128 individually shielded strip-like tRPC cells  distributed in four modules symmetrically placed around the beam axis.

To cope with the expected maximum particle load, the detector is operated above room temperature in order to increase its count rate capability by decreasing the glass resistivity. 

The detector has been successfully operated under a maximum particle rate of about $600$~Hz/cm$^2$ at a temperature of $31.5^\circ$C during a production beam time for six weeks showing an average resolution of about $80$~ps within the requirements of the experiment. 

\section{Acknowledgments}
This work was supported by Fundação para a Ciência e Tecnologia, Portugal, in the framework of a signed MoU with HADES and by the European Union’s Horizon 2020 research and innovation program under grant agreement no. 824093 (STRONG2020).

\end{document}